\documentclass[twocolumn,superscriptaddress,PRL]{revtex4-1}

\usepackage{amsmath}
\usepackage{amssymb}
\usepackage{graphicx}
\usepackage{color}
\usepackage{bm}
\usepackage{enumerate}

\begin{document}

\title{Two distinct superconducting pairing states divided by the nematic end point in FeSe$_{\bm{1-x}}$S$_{\bm{x}}$}

\author{T.~Hanaguri}
\email{hanaguri@riken.jp}
\affiliation{RIKEN Center for Emergent Matter Science, Wako, Saitama 351-0198, Japan}

\author{K.~Iwaya}
\affiliation{RIKEN Center for Emergent Matter Science, Wako, Saitama 351-0198, Japan}

\author{Y.~Kohsaka}
\affiliation{RIKEN Center for Emergent Matter Science, Wako, Saitama 351-0198, Japan}

\author{T.~Machida}
\affiliation{RIKEN Center for Emergent Matter Science, Wako, Saitama 351-0198, Japan}

\author{T.~Watashige}
\affiliation{Department of Physics, Kyoto University, Kyoto 606-8502, Japan}

\author{S.~Kasahara}
\affiliation{Department of Physics, Kyoto University, Kyoto 606-8502, Japan}

\author{T.~Shibauchi}
\affiliation{Department of Advanced Materials Science, University of Tokyo, Chiba 277-8561, Japan}

\author{Y.~Matsuda}
\affiliation{Department of Physics, Kyoto University, Kyoto 606-8502, Japan}

\date{\today}

\begin{abstract}
{
Unconventional superconductivity often competes or coexists with other electronic orders.
In iron-based superconductors, relationship between superconductivity and the nematic state, where the lattice rotational symmetry is spontaneously broken in the electronic states, has been discussed but unclear.
Using spectroscopic-imaging scanning tunneling microscopy, we investigate how the band structure and the superconducting gap evolve in FeSe$_{1-x}$S$_x$, as the sulfur substitution suppresses nematicity that eventually diminishes at the nematic end point (NEP) at $x=0.17$.
Anisotropic quasiparticle-interference patterns, which represent the nematic band structure, gradually become isotropic with increasing $x$ without detectable anomalies in the band parameters at the NEP.
By contrast, the superconducting gap, which is almost intact in the nematic phase, suddenly shrinks as soon as $x$ exceeds the NEP.
Our observation implies that the presence or absence of nematicity results in two distinct pairing states, whereas the pairing interaction is insensitive to the strength of nematicity.
This provides a clue for understanding the pairing mechanism.
}
\end{abstract}

\maketitle

Iron-based superconductors are characterized by multiple Fermi surface pockets, each of which is composed of multiple Fe 3$d$ orbitals~\cite{Singh2008PRL,Kuroki2008PRL,Graser2009NJP}.
Such a complicated electronic structure leads to spin~\cite{Kuroki2008PRL,Mazin2008PRL,Hirschfeld2011ROPP} and orbital~\cite{Kontani2010PRL,Kontani2012SSC,Yanagi2010PRB,Yanagi2010JPSJ} fluctuations that are major candidates for the pairing interaction causing superconductivity.
Electron interactions responsible for the pairing often give rise to other ordered phases that compete or coexist with superconductivity.
In iron-based superconductors, the nematic phase, which is characterized by the broken lattice rotational symmetry in the electronic states, is regarded as a sister phase of superconductivity~\cite{Fernandes2014NatPhys}.
Therefore, the interplay between superconductivity and nematicity is the key to clarifying the nature of the dominant fluctuation responsible for superconductivity.

Iron chalcogenide solid solution FeSe$_{1-x}$S$_x$ is an important model system for investigating the relationship between superconductivity and nematicity.
The parent material FeSe exhibits tetragonal-orthorhombic structural phase transition at $T_{\rm s} \sim 90$~K, without a subsequent magnetic order at a lower temperature~\cite{Hsu2008PNAS,Mizuguchi2010JPSJ}.
Angle-resolved photoemission spectroscopy (ARPES) experiments have revealed that the structural transition accompanies the lifting of the degeneracy between $d_{xz}$ and $d_{yz}$ orbitals, of which energy splitting reaches values as large as $\sim50$~meV at low temperatures~\cite{Shimojima2014PRB,Nakayama2014PRL,Watson2015PRB}.
Although details of the interpretation of the ARPES data are still under debate~\cite{Suzuki2015PRB,Watson2016PRB,Fedorov2016SciRep}, it is widely accepted that such a large energy splitting is difficult to be explained by the small orthorhombic lattice distortion ($\sim0.2$~\%)~\cite{Yamakawa2016PRX}, meaning that the orthorhombic phase is not associated with the lattice instability but nothing other than the electronically-driven nematic phase.
Superconductivity takes place in this nematic phase at $T_{\rm c} \sim 9$~K~\cite{Hsu2008PNAS}.
Nematicity can be systematically suppressed by isovalent sulfur substitution and $T_{\rm s}$ diminishes at the NEP at $x=0.17$~\cite{Mizuguchi2010JPSJ,Hosoi2016PNAS}, whereas superconductivity is observed all the way up to $x=1.0$~\cite{Mizuguchi2009JPSJ,Xing2016PRB}.
High-quality single crystals are available over a wide $x$ range including the NEP~\cite{Hosoi2016PNAS}, offering an ideal platform for experimentally examining the electronic-state evolutions as a function of $x$, namely the strength of nematicity.

There are several experiments that have examined the electronic states of FeSe$_{1-x}$S$_x$.
ARPES~\cite{Watson2015PRB2,Reiss2017PRB} and quantum-oscillations~\cite{Coldea2016arXiv} experiments have revealed that the size of the Fermi surface increases with increasing $x$ and in-plane band anisotropy disappears in the tetragonal phase~\cite{Reiss2017PRB}.
Superconducting (SC) gap has been detected by ARPES~\cite{Xu2016PRL} and scanning tunneling microscopy (STM)~\cite{Moore2015PRB}, but $x$ ranges so far studied are limited.
Comparative studies of the band structure, and the SC gap across the NEP have remained unexplored.

To settle this issue, we have performed spectroscopic-imaging STM (SI-STM) on FeSe$_{1-x}$S$_x$ single crystals over a wide $x$ range $0 \leq x \leq 0.25$.
Topographic and spectroscopic images gained by SI-STM allow us to concurrently determine the chemical composition, the band structure and the SC gap in the same field of view.
By repeating the same measurements and analyses on the samples with different $x$, we have examined the systematic evolutions of the electronic states across the NEP and have provided a spectroscopic evidence of phase transition in the SC state at the NEP.

\subsection*{Determining the sulfur contents by topographic imaging}

First we characterize the samples by topographic imaging.
Figure~1a-e show the representative atomic-resolution topographic images of the cleaved surfaces of FeSe$_{1-x}$S$_x$ single crystals.
\begin{figure*}[tb]
\includegraphics[width=150mm]{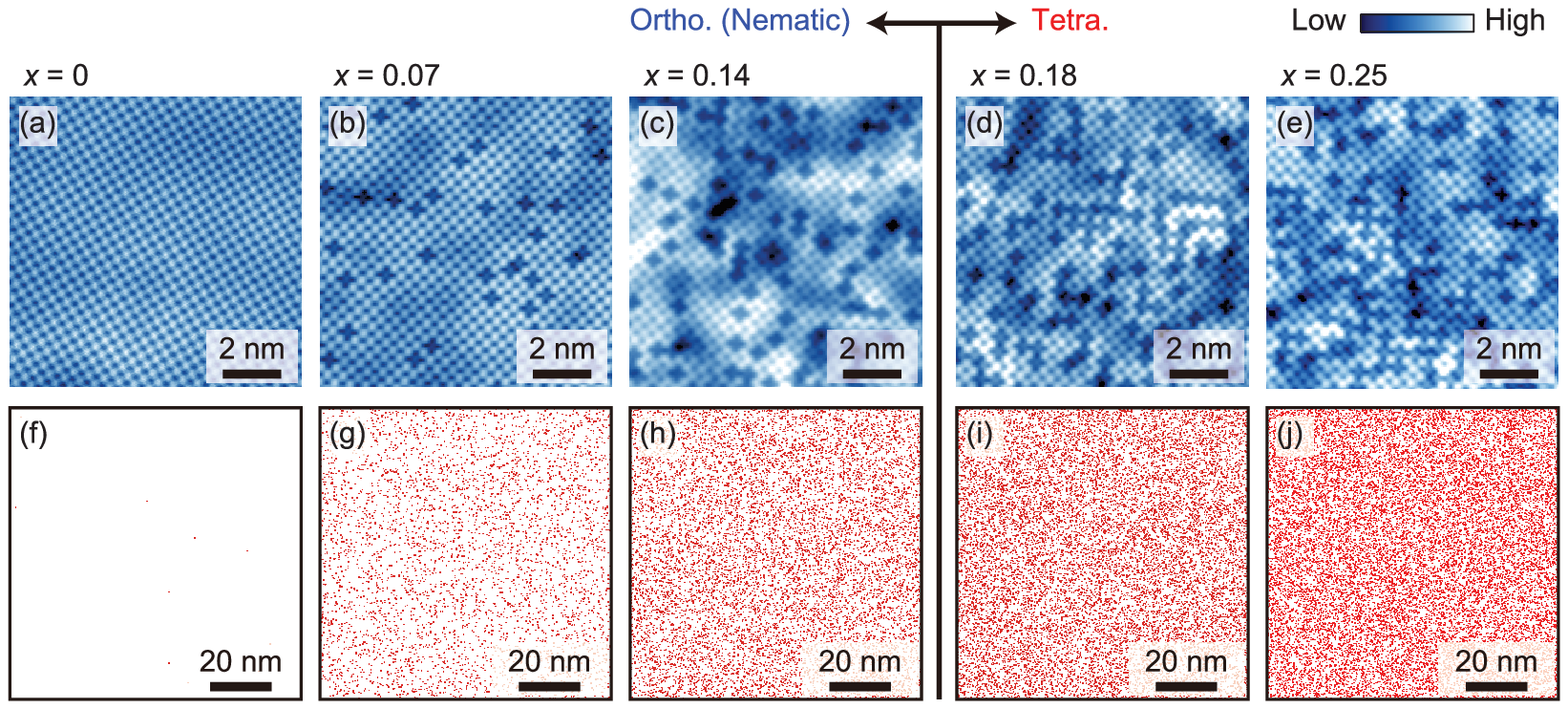}
\caption{
\textbf{Characterization of FeSe$_{\bm{1-x}}$S$_{\bm{x}}$ single crystals by topographic imaging.}
\textbf{a-e},
Constant-current topographic STM images of FeSe$_{1-x}$S$_x$.
Feedback loop for scanning was established at the setup sample-bias voltage $V_{\rm s}=+20$~mV and setup tunneling current $I_{\rm s}=100$~pA.
For $x=0$ sample, $I_{\rm s}=10$~pA was used.
Local depressions represent sulfur atoms that replace the selenium atoms.
The images were cropped from the larger images of 100~nm $\times$ 100~nm (160~nm $\times$ 160~nm for $x=0$) fields of view, where all the subsequent SI-STM experiments were performed.
\textbf{f-j},
Distributions of sulfur atoms in wider fields of view.
Each red dot denotes the position of a sulfur atom.
No tendency to the segregation is observed.
}
\end{figure*}
In the sulfur-substituted samples, depressions are observed in a regular chalcogen lattice.
These correspond to the sulfur atoms that have a smaller atomic radius than selenium.
By counting the numbers of sulfur and selenium atoms in a wide field of view (100~nm $\times$ 100~nm), the chemical composition can be determined with high accuracy.
In this study, we have measured 9 different samples ranging $0 \leq x \leq 0.25$.
Next we checked the spatial distribution of the sulfur atoms by plotting the locations of the sulfur atoms.
As shown in Fig.~1f-j, the sulfur atoms are uniformly distributed in our fields of view without apparent tendency to the segregation.
Details of the sample characterization by topographic imaging are given in Supplementary Information section I.

\subsection*{Evolution of the band structure upon sulfur doping}

Having the chemical compositions of all the samples, we examine the systematic $x$ evolution of the band structure by analyzing the electronic standing waves generated by the quasiparticle interference (QPI) effect.
We observe the QPI patterns that appear in the energy ($E$)-dependent tunneling-conductance map $g(\mathbf{r},E=eV)\equiv dI(\mathbf{r},V)/dV$, where $\mathbf{r}$ is the lateral position at the surface, $e$ is the elementary charge, $V$ is the sample bias voltage, and $I$ is the tunneling current.
In this study, we analyze the normalized conductance map $L(\mathbf{r},E=eV)\equiv (dI(\mathbf{r},V)/dV)/(I(\mathbf{r},V)/V)$ to avoid the error associated with the feed-back loop of the constant-current scanning~\cite{Kohsaka2007Science,Feenstra1987SurfSci}.
Fourier transformation of $L(\mathbf{r},E)$ yields QPI patterns $L_q(\mathbf{q},E)$ in scattering-vector $\mathbf{q}$ space, allowing us to determine the set of characteristic scattering wave vectors $\mathbf{q}_i(E)$ that connect two states on the same constant-energy surface in momentum $\mathbf{k}$ space.
The band dispersions can be inferred from $\mathbf{q}_i(E)$.

In order to study the normal-state characters down to low energies, we suppressed the SC gap by applying a magnetic field of 12~T perpendicular to the surface.
As we will explain later, the SC gap shrinks above the NEP.
Therefore, we analyzed the zero-field data in the tetragonal phase.
Figure~2 shows $x$-dependent $L_q(\mathbf{q},E)$ at representative energies.
In FeSe ($x=0$), QPI patterns exhibit strong in-plane anisotropy, manifesting itself as the nematic band structure.
\begin{figure*}[tb]
\includegraphics[width=150mm]{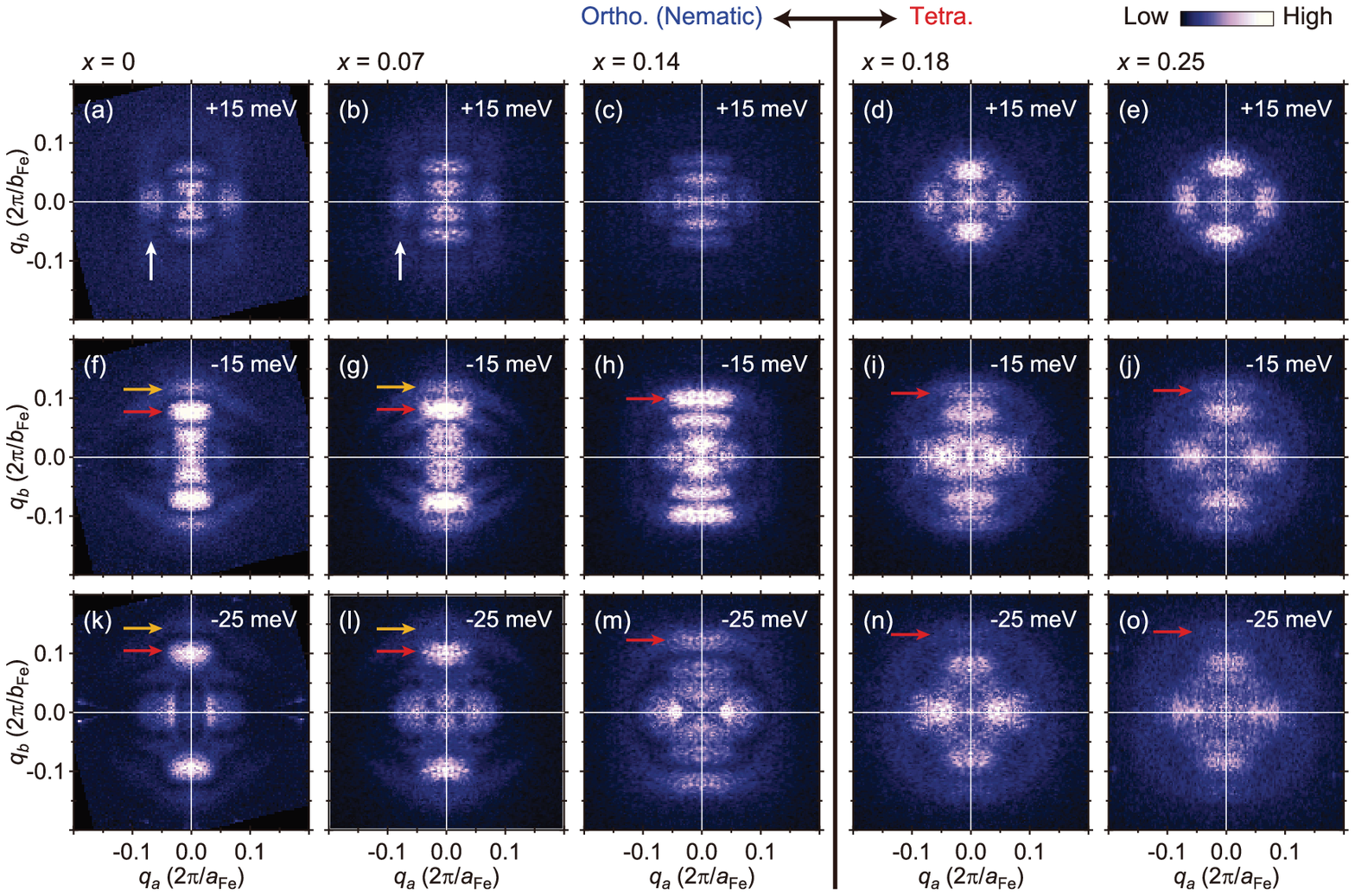}
\caption{
\textbf{In-plane QPI patterns of FeSe$_{\bm{1-x}}$S$_{\bm{x}}$ at representative energies.}
Each panel depicts the QPI pattern $L_q(\mathbf{q},E)$ in $\mathbf{q}$ space obtained by Fourier transforming the normalized real-space conductance map $L(\mathbf{r},E)$.
Tunneling spectra for each $L(\mathbf{r},E)$ map were taken on a grid of 256 $\times$ 256 pixels in a 100~nm $\times$ 100~nm (160~nm $\times$ 160~nm for $x=0$) field of view.
Measurement conditions were $V_{\rm s}=+50$~mV, $I_{\rm s}=100$~pA and the modulation amplitude for the lock-in detection $V_{\rm mod} = 0.85$~mV$_{\rm rms}$.
For the samples in the nematic phase ($x<0.17$), experiments were done under 12~T magnetic field perpendicular to the surfaces to suppress superconductivity.
In other samples, no magnetic field was applied.
Strongly anisotropic QPI patterns for the $x=0$ sample (\textbf{a},\textbf{f} and \textbf{k}) reflect the nematic character of the band structure.
White allow indicates the inter-electron-band QPI signal.
Red and orange allows denote the QPI signals associated with the outer hole band (see text for details).
The QPI patterns gradually become isotropic with increasing $x$.
}
\end{figure*}
In the filled state ($E<0$), there are spot- and arc-like features (red and orange arrows, respectively) along the $\mathbf{q}_b$ axis whereas in the empty state ($E>0$), a streak-like feature (white arrow) appears perpendicular to the $\mathbf{q}_a$ axis.
Here, $\mathbf{q}_a \parallel \mathbf{a}_{\rm Fe}$ and $\mathbf{q}_b \parallel \mathbf{b}_{\rm Fe}$ where $\mathbf{a}_{\rm Fe}$\ and $\mathbf{b}_{\rm Fe}$ are in-plane primitive vectors of the Fe lattice.
In this article, we adopt a coordinate system with $\mathbf{x} \parallel \mathbf{a_{\rm Fe}}$, $\mathbf{y} \parallel \mathbf{b_{\rm Fe}}$ and $b_{\rm Fe} \geq a_{\rm Fe}$.
As we reported earlier~\cite{Kasahara2014PNAS} and will be discussed in detail below, these QPI signals exhibit electron- and hole-like dispersions along $\mathbf{q}_a$ and $\mathbf{q}_b$, respectively, corresponding to the intra-band back scatterings in the electron bands at the Brillouin zone corner and in the hole band at the zone center, respectively.
With increasing $x$, spot-like features gradually spread out in $\mathbf{q}$ space and evolve into almost isotropic QPI patterns above the NEP, as expected from the tetragonal symmetry.
Residual anisotropy observed above the NEP may be associated with an unavoidable small strain in the sample that could matter because of the large nematic susceptibility~\cite{Hosoi2016PNAS}.
Or it would be intrinsic in origin as suggested in other tetragonal iron-based superconductors~\cite{Kasahara2012Nature,Singh2015SciAdv,Thewalt2017arXiv}.
A QPI data set including results of other $x$ and $E$ values is available in Supplementary Information section II.

Energy-dependent line profiles from Fig.~2 along the $\mathbf{q}_a$ and $\mathbf{q}_b$ axes are shown in Fig.~3a-e and Fig.~3f-j, respectively.
\begin{figure*}[tb]
\includegraphics[width=150mm]{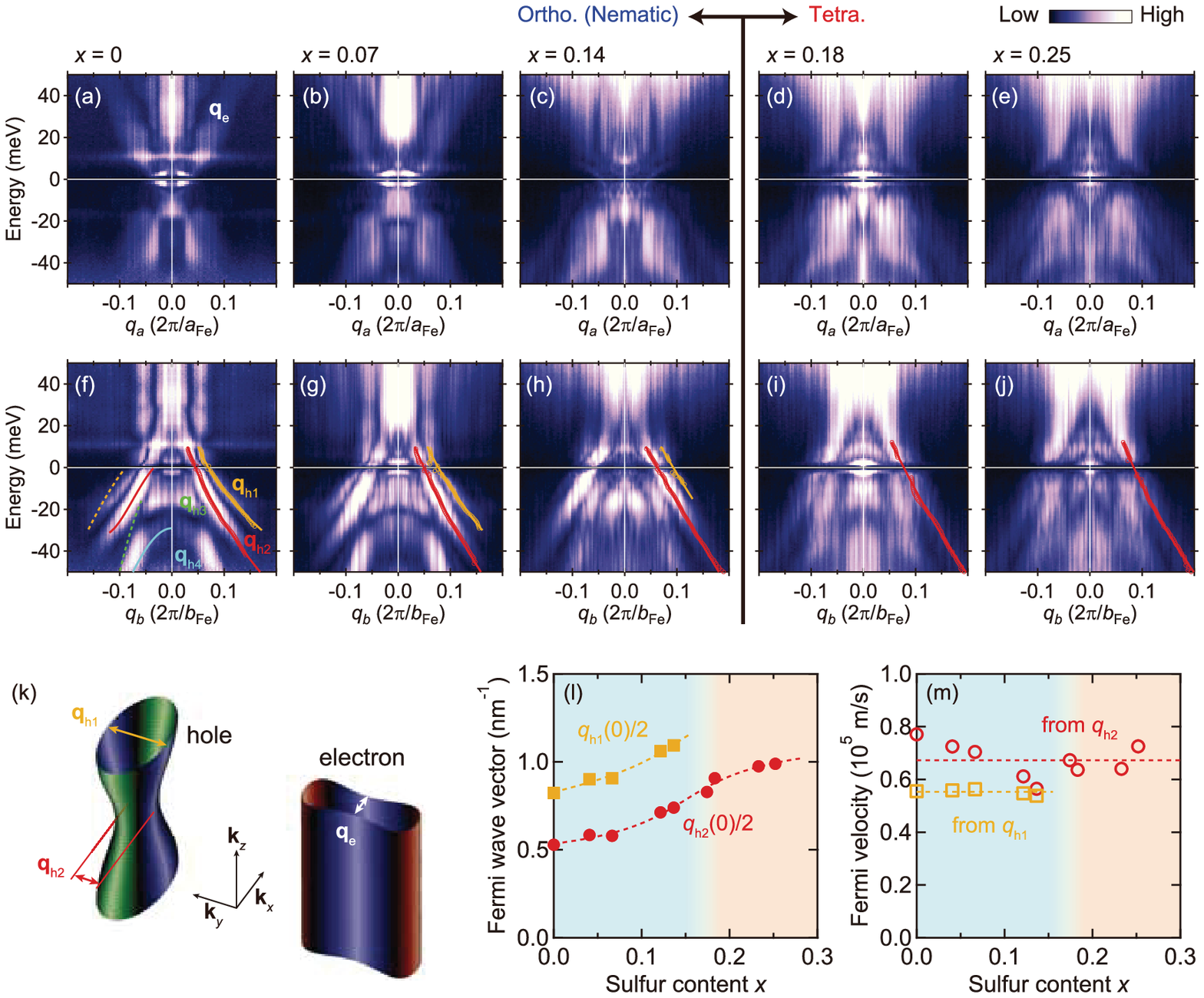}
\caption{
\textbf{Evolutions of QPI dispersion and band structure upon sulfur doping.}
\textbf{a-e},
Energy-dependent line profiles of Fig.~2 along the $\mathbf{q}_a$ direction.
Electron-like QPI branch $\mathbf{q}_e$ is identified.
This branch becomes obscured with increasing $x$.
\textbf{f-j},
Energy-dependent line profiles of Fig.~2 along the $\mathbf{q}_b$ direction.
There are four QPI branches as labeled $\mathbf{q}_{{\rm h}i}$ ($i=1-4$).
In order to analyze the dispersions of QPI branches that cross the Fermi level ($\mathbf{q}_{{\rm h}1}$ and $\mathbf{q}_{{\rm h}2}$), we have determined the peak positions of the constant-energy line profile (scattering-vector distribution curve) by fitting the data with multiple Gaussian peaks.
The obtained peak positions for $\mathbf{q}_{{\rm h}1}$ and $\mathbf{q}_{{\rm h}2}$ are plotted by orange and red open circles, respectively, on the right half of each panel.
The obtained $\mathbf{q}_{{\rm h}1}(E)$ and $\mathbf{q}_{{\rm h}2}(E)$ data are fitted by quadratic functions of $E$ shown by orange and red solid lines on the right half of each panel.
On the left half of \textbf{f}, intra-hole-band backscattering vectors expected from the ARPES data~\cite{Watson2015PRB} are plotted.
Orange and red lines are for the outer hole band, and green and light blue lines are for the inner hole band.
Solid and dashed lines represents the back scattering $\mathbf{q}$ vectors at $k_z=0$ and $k_z=\pi/c$, respectively.
\textbf{k},
Schematic illustration of the constant energy surface in $\mathbf{k}$ space (not in scale).
Colors on the surface denote different orbital characters;
blue: $d_{xz}$,
green: $d_{yz}$,
red: $d_{xy}$.
The inner hole band is not shown.
\textbf{l},
Evolutions of the Fermi wave vectors as functions of sulfur content $x$.
$q_{\rm {h}1}(E = 0 {\rm ~meV})/2$ and $q_{\rm {h}2}(E = 0 {\rm ~meV})/2$ correspond to the minor axes of the elliptical cross sections of the outer hole Fermi cylinder at $k_z=\pi/c$ and $k_z=0$, respectively.
Lines are the guides for the eyes.
\textbf{m},
Evolution of the Fermi velocity as a function of sulfur content $x$.
Lines are the guides for the eyes.
Anomalies are observed neither in the Fermi wave vector nor in the Fermi velocity at the NEP at $x=0.17$.
}
\end{figure*}
In the $x=0$ sample, there is one electron-like branch ($\mathbf{q}_{\rm e}$) along $\mathbf{q}_a$ (Fig.~3a), which we ascribe to the intra-electron-band scattering that connects the flat parts of the constant-energy surface schematically shown in Fig.~3k.
Features appearing along $\mathbf{q}_b$ are rather complicated.
There are four $\mathbf{q}$ vectors labeled as $\mathbf{q}_{{\rm h}i}$ ($i=1-4$) in Fig.~3f.
Two of them, $\mathbf{q}_{\rm h1}$ and $\mathbf{q}_{\rm h2}$, cross the Fermi level ($E=0$~meV).
Since all of these vectors exhibit hole-like dispersions, they should correspond to the intra-hole-band scatterings.
To clarify the relation between these QPI branches and the band structure, we compare our QPI data with the $\mathbf{q}$ vectors expected from the intra-band back scatterings ($\mathbf{k} \leftrightarrow -\mathbf{k}$) in the hole bands.
According to the ARPES results~\cite{Watson2015PRB,Maletz2014PRB}, there are two concentric hole bands and the outer one crosses the Fermi level.
In contrast to the strongly two-dimensional electron band~\cite{Watson2016PRB}, the hole bands possess three dimensionality and the constant-energy surface is warping along the $\mathbf{k}_z$ direction (Fig.~3k).
Because of this warping, back scattering $\mathbf{q}$ vectors at $k_z=0$ and $k_z=\pi/c$ are different.
($c$ denotes the lattice constant along the $\mathbf{z}$ axis.)
We calculate various back scattering $\mathbf{q}$ vectors from the ARPES results~\cite{Watson2015PRB} and plot them in the left half of Fig.~3f.
Here, orange and red lines are for the outer hole band, and green and light blue lines are for the inner hole band.
Solid and dashed lines represents the back scattering $\mathbf{q}$ vectors at $k_z=0$ and $k_z=\pi/c$, respectively.
QPI branches crossing the Fermi energy, $\mathbf{q}_{\rm h1}$ and $\mathbf{q}_{\rm h2}$ coincide with the back scatterings in the outer hole band at $k_z=\pi/c$ and $k_z=0$, respectively, meaning that they measure the minor axes of the elliptical cross sections of the cylindrical constant-energy surface (Fig.~3k).
Here we note that $\mathbf{q}_{\rm e}$, $\mathbf{q}_{\rm h1}$ and $\mathbf{q}_{\rm h2}$ all connect the parts of the constant-energy surface with a dominant $d_{xz}$ character~\cite{Suzuki2015PRB,Sprau2017Science,Kreisel2017PRB}.
Other branches $\mathbf{q}_{\rm h3}$ and $\mathbf{q}_{\rm h4}$ should be associated with the inner hole band because they are close to the corresponding back scattering $\mathbf{q}$ vectors at $k_z=\pi/c$ and $k_z=0$, respectively (Fig.~3f).

The above comparison between QPI and ARPES data indicates that the different $k_z$ states of the hole bands are separately captured in the QPI patterns.
This provides a new perspective because most of QPI studies to date either have been performed on two-dimensional materials or have implicitly assumed the two-dimensional electronic states.
The microscopic origin of this three-dimensional QPI is unclear at present.
One possible scenario is that, extrema in the constant-energy surface that bring about large density of states may play a role, as in the case of the quantum oscillations.
In particular, if there is a flat part in the three-dimensional constant-energy surface, strong QPI signals perpendicular to the flat part may show up due to electron focusing~\cite{Weismann2009Science}.
Detailed theoretical analyses are indispensable to establish the condition to observe the three-dimensional QPI.

Besides $\mathbf{q}_{\rm e}$ and $\mathbf{q}_{{\rm h}i}$ ($i=1-4$), there are a couple of features in Fig.~3a,f but they may not be related to the band structure.
Features near $|\mathbf{q}|\sim 0.05(2\pi/a_{\rm Fe})$ in the filled and empty states along $\mathbf{q}_a$ and $\mathbf{q}_b$, respectively, do not disperse with energy, being inconsistent with any band dispersions.
Signals near $\mathbf{q} \sim 0$ may be due to the random distribution of the defects.

Having these assignments of the QPI branches, we examine the $x$ dependence of the band structure.
As shown in Fig.~3a-j, the electron-like branch becomes obscured with increasing $x$ whereas the hole-like branches remain observable over the whole $x$ range studied.
For the outer hole band, the relative difference between $\mathbf{q}_{\rm h1}$ and $\mathbf{q}_{\rm h2}$ becomes smaller with increasing $x$, and the $\mathbf{q}_{\rm h1}$ branch is not clearly resolved for $x>0.14$, suggesting that three-dimensionality becomes weaker with increasing $x$.

To analyze the evolution of the band structure quantitatively, we fit dispersions of $\mathbf{q}_{\rm h1}$ and $\mathbf{q}_{\rm h2}$ (right halves of Fig.~3f-j), from which Fermi wave vectors $q_{{\rm h}i}(E=0{\rm ~meV})/2$ and Fermi velocities $(2/\hbar) dE/dq_{{\rm h}i}|_{E=0{\rm ~meV}}$ can be calculated ($i = 1, 2$).
As shown in Fig.~3l, Fermi wave vectors, which correspond to the minor axes of the cross sections of warped Fermi cylinder at $k_z=0$ and $k_z=\pi/c$, increase with increasing $x$.
Since the increasing rates of the cross sectional areas of the Fermi cylinder~\cite{Coldea2016arXiv} are almost the same as the increasing rates of the Fermi wave vectors for the minor axes, it is suggested that the major axes of the cross sections of the Fermi cylinder are hardly affected by sulfur doping, meaning that the in-plane anisotropy decreases with increasing $x$.
Fermi velocities are almost constant or slightly decrease with $x$ as depicted in Fig.~3m.
The important observation is that both Fermi wave vector and Fermi velocity evolve smoothly without detectable anomalies at the NEP.

\subsection*{Evolution of the SC gap upon sulfur doping}

Next we examine the $x$ dependence of the SC gap.
Figure~4a shows a series of SC gap spectra averaged over the same fields of view used for the above QPI experiments.
\begin{figure*}[tb]
\includegraphics[width=150mm]{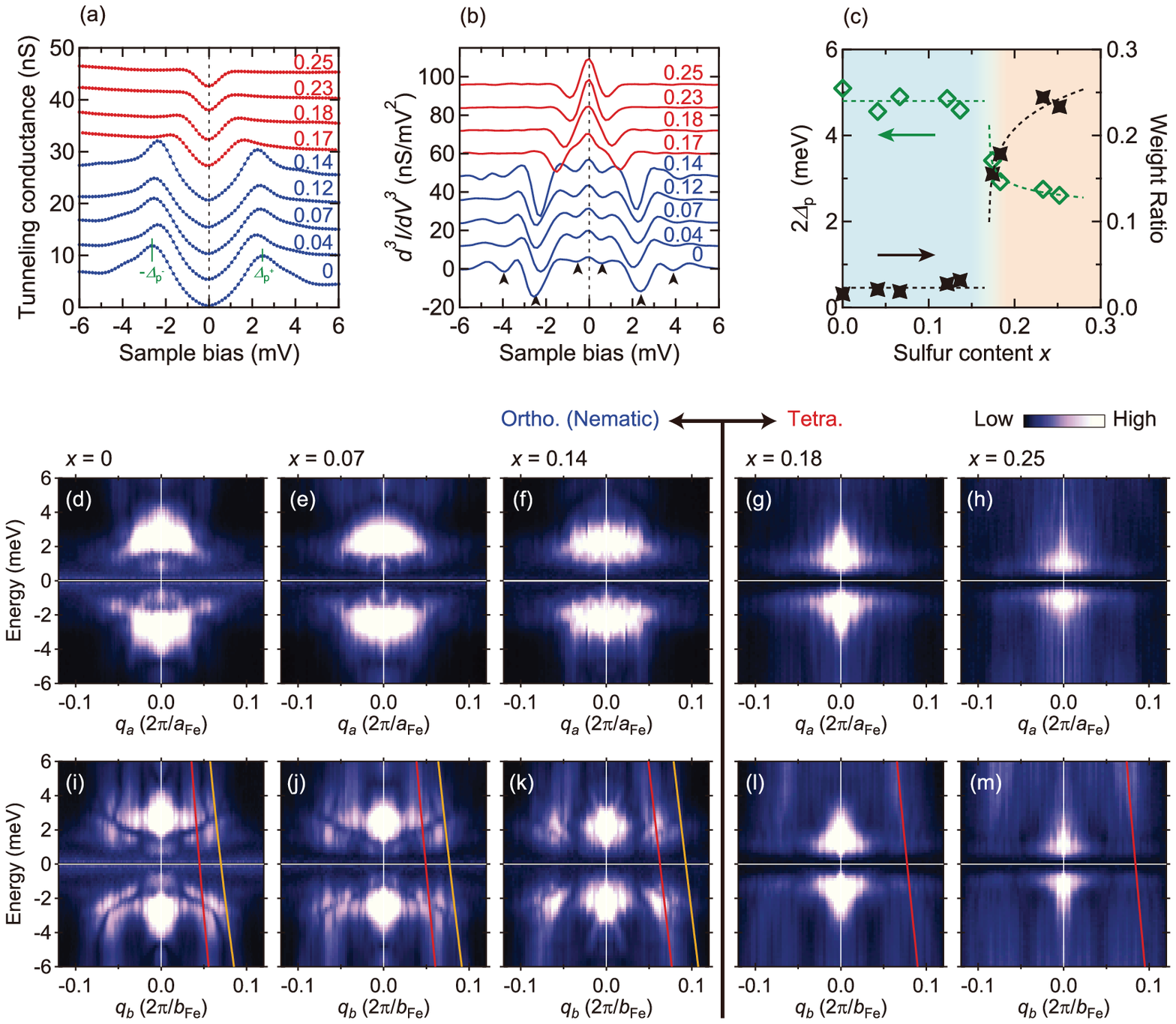}
\caption{
\textbf{Evolutions of SC gap and BQPI pattern upon sulfur doping.}
\textbf{a},
Averaged tunneling spectra of FeSe$_{1-x}$S$_x$ showing the SC gaps.
Each curve is shifted vertically by 5~nS for clarity.
Spectra before the averaging were acquired on a grid of 256$\times$256 pixels in a 100~nm $\times$ 100~nm (160~nm $\times$ 160~nm for $x=0$) field of view.
Measurement conditions were $V_{\rm s}=+20$~mV, $I_{\rm s}=100$~pA and $V_{\rm mod} = 0.21$~mV$_{\rm rms}$.
The same data sets were used to obtain BQPI patterns shown in \textbf{d-m}.
\textbf{b},
Energy second derivative of averaged tunneling spectra shown in \textbf{a}.
Dips marked by arrows correspond to the peak-like features in the original tunneling spectra.
Each curve is shifted vertically by 12~nS/mV$^2$ for clarity.
\textbf{c},
Evolutions of the apparent gap amplitude $\Delta_{\rm p}^{+}+\Delta_{\rm p}^{-}$ and the zero-energy spectral weight normalized by the weights at the gap-edge energies $2g(0)/(g(\Delta_{\rm p}^{+})+g(-\Delta_{\rm p}^{-}))$.
In both quantities, there are abrupt changes at the NEP at $x=0.17$.
Lines are the guides for the eyes.
\textbf{d-h},
Energy-dependent line profiles of low-energy $L_q(\mathbf{q},E)$'s along the $\mathbf{q}_a$ direction showing BQPI patterns.
\textbf{i-m},
Same as \textbf{d-h} but along the $\mathbf{q}_b$ direction.
Orange and red lines denote the fitted dispersions of $\mathbf{q}_{{\rm h}1}(E)$ and $\mathbf{q}_{{\rm h}2}(E)$ shown in Fig.~3\textbf{f-j}.
In the nematic phase, there are two pairs of V-shaped dispersing BQPI branches that are smoothly connected to $\mathbf{q}_{{\rm h}1}(E)$ and $\mathbf{q}_{{\rm h}2}(E)$. These BQPI branches disappear above the NEP at $x=0.17$.
}
\end{figure*}
The spectrum in the nematic phase ($x < 0.17$) is apparently unaffected by sulfur doping.
The V-shaped spectra indicate nodal or highly-anisotropic SC gap~\cite{Song2011Science}.
We take second derivative of the tunneling spectra to examine the fine structures of the gap.
As shown in Fig. 4b, there are dips, which correspond to the peaks in the original spectra, at about $\pm 0.6$~mV, $\pm 2.5$~mV and $\pm 4.0$~mV.
These fine details of the SC gap are intact as long as the system is in the nematic phase.
The situation changes drastically at the NEP, above which the SC gap suddenly shrinks.
In Fig.~4c, we plot apparent gap amplitude $\Delta_{\rm p}^{+}+\Delta_{\rm p}^{-}$ and the zero-energy spectral weight normalized by the weights at the gap-edge energies $2g(0)/(g(\Delta_{\rm p}^{+})+g(-\Delta_{\rm p}^{-}))$.
Here $\Delta_{\rm p}^{+(-)}$ denotes the energy of the main peak in the tunneling spectrum $g(E)$ in the positive (negative) bias side (Fig.~4a).
It is clear that there are abrupt changes in both quantities at the NEP, manifesting the phase transition in the SC state.
Recent specific-heat measurement shows that tetragonal $x=0.20$ sample has $T_{\rm c} \sim 4.5$~K that is lower than $T_{\rm c} = 9 \sim 11$~K in the nematic samples~\cite{Sato2017arXiv}, being consistent with the SC-gap suppression observed in this study.

To obtain the $\mathbf{k}$-space information of this transition, we study QPI patterns of Bogoliubov quasiparticles observed in the SC-gap energy scale.
Energy-dependent line profiles of the Fourier-transformed Bogoliubov QPI (BQPI) patterns along $\mathbf{q}_a$ and $\mathbf{q}_b$ are shown in Fig.~4d-h and Fig.~4i-m, respectively.
In both directions, patterns are almost particle-hole symmetric, being consistent with the nature of the Bogoliubov quasiparticles.
In the $x=0$ sample, there are two pairs of V-shaped dispersing branches along $\mathbf{q}_b$ (Fig.~4i).
The lower- and higher-energy branches terminate at about $\pm 2.5$~mV and $\pm 4.0$~mV, and are connected to the $\mathbf{q}_{\rm h1}$ and $\mathbf{q}_{\rm h2}$ branches, respectively.
This suggests that the V-shaped dispersions are originated from the SC gap on the outer hole band and the dips at $\pm 2.5$~mV and $\pm 4.0$~mV in Fig.~4b represent the gap amplitudes near $k_z=\pi/c$ and $k_z=0$, respectively.
The dip at $\pm 0.6$~mV should also represent a feature related to the SC gap but its location in $\mathbf{k}$ space is unclear at present.
We could not observe BQPI signals associated with the electron band~\cite{Sprau2017Science}.

In general, dispersing BQPI signals mean that the SC gap is anisotropic in $\mathbf{k}$ space.
In such a case, a constant energy surface in $\mathbf{k}$ space surrounds the $\mathbf{k}$ point where gap takes minimum, and the dominant scattering vectors may connect the tips of the constant energy surface that make joint density of states high.
This model was developed in the field of cuprate superconductivity~\cite{Hoffman2002Science,McElroy2003Nature,Wang2003PRB} and has also been applied to other unconventional superconductors including FeSe~\cite{Sprau2017Science}.
In this way, dispersing BQPI signals carry information of the gap structure in $\mathbf{k}$ space.
In the case of FeSe, if the observed V-shaped dispersions along $\mathbf{q}_b$ are associated with the intra-outer-hole-band back scattering, the gap minimum should be located along the major axis of the elliptical cross section of the Fermi surface.
We could not reconstruct the whole structure of the SC gap because the corresponding dispersions along $\mathbf{q}_a$ are not clear in our data (Fig.~4d).
Nevertheless, the inferred in-plane gap anisotropy is consistent with the previous ARPES~\cite{Xu2016PRL} and SI-STM~\cite{Sprau2017Science} results.
Moreover, the observed two Bogoliubov QPI branches along $\mathbf{q}_b$ that come from $k_z=0$ and $k_z=\pi/c$ states exhibit similar V-shaped dispersions, suggesting that the in-plane gap anisotropy does not depend on $k_z$.

Sulfur doping does not alter the two pairs of V-shaped Bogoliubov QPI branches as long as the system is in the nematic phase (Fig.~4i-k).
Above the NEP, however, both branches disappear.
This is in accord with the disappearance of the $\pm 2.5$~mV and $\pm 4.0$~mV features in the averaged tunneling spectra (Fig.~4a,b), confirming that they are associated with the outer hole band.
The $\mathbf{k}$-space structure of the SC gap in the tetragonal phase is unclear at present.
Recent thermal conductivity measurement suggests that strong anisotropy in the SC gap remains in the tetragonal phase~\cite{Sato2017arXiv}.
It is an interesting future subject to perform higher energy-resolution BQPI experiments at lower temperatures to clarify the details of the $\mathbf{k}$-space structure of the SC gap in the tetragonal phase.

\subsection*{Discussion}

Our concurrent study of the band structure and the SC gap in FeSe$_{1-x}$S$_x$ has revealed novel aspects of the interplay between nematicity and superconductivity.
Upon sulfur substitution, the band anisotropy diminishes.
This process evolves smoothly over the $x$ range studied; no anomaly has been detected at the NEP.
Within the nematic phase, the Fermi wave vector changes by a several tens of percent.
In sharp contrast, even fine details of the SC gap are intact in the nematic phase.

The most striking observation is that there is a phase transition in the SC state at the NEP that accompanies a significant suppression of the SC gap on the outer hole band in the tetragonal phase.
This should not be a result of enhanced quasiparticle damping, which could be caused by the strong nematic fluctuations, because residual resistivity does not exhibit noticeable increase above the NEP~\cite{Hosoi2016PNAS}.
Rather, it is reasonable to assume that the presence or absence of nematicity results in two distinct pairing states.
Since the nematic phase is characterized by the lifting of degeneracy between $d_{xz}$ and ${d_{yz}}$ orbitals, it is suggested that the pairing interaction works effectively only on one of these two orbitals, but not on both of them simultaneously.
This is consistent with the orbital-selective pairing scenario suggested by the analysis of the SC gap structure in $\mathbf{k}$ space~\cite{Sprau2017Science}.
The insensitivity of the SC gap to the band structure in the nematic phase indicates that, as soon as the nematicity sets in, the pairing interaction grows rapidly and saturates immediately.

It is important to compare the effect of isovalent sulfur substitution with that of hydrostatic pressure.
As sulfur has smaller atomic radius than selenium, sulfur substitution may cause chemical pressure that mimics the application of hydrostatic pressure.
Indeed, the suppression of the nematic transition temperature can commonly be initiated by sulfur substitution~\cite{Mizuguchi2010JPSJ,Hosoi2016PNAS} and by external hydrostatic pressure~\cite{Medvedev2009NatMat,Sun2016NatCommun}.
However, being in sharp contrast to the suppressed SC gap in FeSe$_{1-x}$S$_x$ above the NEP, $T_{\rm c}$ is largely enhanced under hydrostatic pressure and reaches about 40~K at $\sim 6$~GPa~\cite{Mizuguchi2010JPSJ,Medvedev2009NatMat,Sun2016NatCommun}.
The key phenomenological difference between sulfur substitution and hydrostatic pressure is that while the former keeps the system nonmagnetic all the way down, the latter induces a magnetic phase~\cite{Sun2016NatCommun,Matsuura2017arXiv}.
The highest $T_{\rm c} \sim 40$~K under hydrostatic pressure is achieved at the verge of the magnetic phase~\cite{Sun2016NatCommun,Matsuura2017arXiv}, suggesting an important role of spin degrees of freedom for the pairing interaction.
In the case of FeSe$_{1-x}$S$_x$, however, it is unlikely that the observed phase transition in the SC state at the NEP is related to the spin degrees of freedom because both nematic and tetragonal phases are nonmagnetic.
Rather, it should predominantly be associated with the imbalance between $d_{xz}$ and ${d_{yz}}$ orbitals as mentioned above.
Namely, pairing interaction in FeSe should be related to both spin and orbital degrees of freedom.
Interplay between spin and orbital may also play a role~\cite{Yamakawa2016PRX}.

Another interesting observation in this work is that, our QPI data separately capture the electronic states at different $k_z$'s.
Although the condition to observe such three-dimensional QPI remains elusive, this may expand the application possibilities of SI-STM technique.
In the case of FeSe$_{1-x}$S$_x$, we find that the SC gap on the outer hole hand near $k_z=0$ is larger than that near $k_z=\pi/c$, whereas the in-plane gap-anisotropy pattern is similar at both $k_z$'s.
We anticipate that these results can act as a touchstone for the theories of iron-based superconductivity.

\subsection*{Methods}

FeSe$_{1-x}$S$_x$ single crystals were prepared by chemical vapor transport technique using KCl and AlCl$_3$ as the transport agents~\cite{Hosoi2016PNAS,Bohmer2013PRB}.
SI-STM experiments were carried out using a commercial low-temperature ultra-high vacuum STM system (Unisoku USM-1300), of which STM unit was replaced with our home-made one~\cite{Hanaguri2006JPCS}.
Clean and flat (001) surfaces necessary for SI-STM were prepared by \textit{in-situ} cleaving at liquid nitrogen temperature under ultra-high vacuum ($\sim 10^{-10}$~Torr).
After the cleaving, samples were immediately transfered to the STM unit kept below $\sim 10$~K.
Electrochemically etched tungsten wires were used as scanning tips.
All tips were cleaned by field evaporation with a field-ion microscope and were checked for the spectra on clean Au(111) or Cu(111) surfaces.
Tunneling spectra were taken with a built-in numerical lock-in detector of a commercial STM controller (Nanonis) at a modulation frequency of 617.3~Hz.
All the SI-STM experiments were performed at 1.5~K, except for the sample with $x=0.23$, which was measured at 0.45~K.
Spectra of the $x=0.23$ sample shown in Fig.~4a,b are numerically convoluted with the Fermi-Dirac function at 1.5~K to compare them with the results of other samples.
Fourier-transformed spectroscopic images were symmetrized by assuming the point group symmetry of $2mm$ to enhance the signal-to-noise ratio.

\subsection*{Data availability}

The data that support the findings of this study are available from the corresponding author upon reasonable request.

\subsection*{Acknowledgments}

The authors thank A. I. Coldea, I. Eremin, H. Kontani, M. D. Watson, and Y. Yamakawa for valuable discussions and comments.
They also appreciate S. Sucharitakul for critical reading of the manuscript.
This work was supported by Grants-in-Aid for Scientific Research from the Ministry of Education, Culture, Sports, Science and Technology of Japan (Grant No. 25220710, No. 15H02106, No. 15H03688, and No. 16H04024).

\subsection*{Author contributions}

T.H. carried out SI-STM experiments and data analyses with assistances from K.I., Y.K., T.M., and T.W.
FeSe$_{1-x}$S$_x$ single crystals were grown by T.W. and S.K.
T.H., T.S. and Y.M. designed and supervised the project.
T.H. wrote the manuscript.
All authors discussed the results and contributed to finalize the manuscript.

\subsection*{Competing financial interests}

The authors declare no competing financial interests.

\end{document}